\documentclass[aps,prl,twocolumn,superscriptaddress,showpacs,longbibliography]{revtex4-1}
\usepackage{graphicx,amsbsy,amsmath,epsfig,natbib,float}
\usepackage{amssymb,latexsym,wasysym,pifont,mathrsfs}
\usepackage{comment,verbatim,multirow,array,sidecap,color}
\usepackage{lineno,setspace,soul,cancel,enumerate,amsmath,mathtools}
\usepackage[normalem]{ulem}
\usepackage[percent]{overpic}

\begin{document}




\title{Avalanches and Structural Change in Cyclically Sheared Silica Glass}

\author{Himangsu Bhaumik} \affiliation{Jawaharlal
  Nehru Center for Advanced Scientific Research, Jakkur Campus,
  Bengaluru 560064, India.}

\author{Giuseppe Foffi} \affiliation{Universit\'e Paris-Saclay, CNRS, Laboratoire de Physique des Solides, 91405 Orsay, France}
\author{Srikanth Sastry} \email{sastry@jncasr.ac.in} \affiliation{Jawaharlal
  Nehru Center for Advanced Scientific Research, Jakkur Campus, Bengaluru 560064, India.}

 
\begin{abstract}  
We investigate avalanches associated with plastic rearrangements and the nature of structural change in the prototypical strong glass, silica, computationally. Although qualitative aspects of yielding in silica are similar to other glasses, we find that the statistics of avalanches exhibits non-trivial behaviour. Investigating the statistics of avalanches and clusters in detail, we propose and verify a new relation between exponents characterizing the size distribution of avalanches and clusters. Across the yielding transition, anomalous structural change and densification, associated with a suppression of tetrahedral order, is observed to accompany strain localisation.

\end{abstract}

\maketitle 



The mechanical response of amorphous solids such as metallic glasses, window glass, foams, emulsions, colloidal suspension {\it etc.}, to external deformation or applied stress is of central importance to characterise their behaviour and determining their utility \cite{Bonn2017c,Nicolas2018,Parmerminireview}. The response for large enough deformations involves plastic rearrangements, leading eventually to yielding. The yielding transition in amorphous solids has been investigated actively in recent years through experiments \cite{SunPRL10,KeimPRL11,AntonagliaPRL14,KeimPRR20,BenninMACROMOL2020}, numerical simulations \cite{maloneyPRE06,karmakarpre10,JaiswalPRL16,Fiocco2013,Priezjev2013,regev2015reversibility,leishangthemNAT2017,Jin,ozawapnas2018,parmarprx19,BarbotPRE20} and theoretical investigations including analysis of elasto-plastic and other models \cite{Dasgupta2012,LinPNAS14,ParisiPNAS17,Urbani2017b,BudrikisNATCOM2017,Popovic2018a,BarlowPRL20,liu2020oscillatory,sastryPRL20,khirallahPRL20,mungan2021metastability}. Yielding has been observed to be a discontinuous transition for sufficiently well annealed glasses under uniform shear \cite{ozawapnas2018} and for cyclic shear \cite{leishangthemNAT2017,parmarprx19,BhaumikPNAS21}, accompanied by a discontinuous drop in energy and stress,  and by localisation of strain in shear bands   \cite{shi2005strain,MartensSM12,RadhakrishnanPRL16,parmarprx19}. 

Plasticity in amorphous solids is distinguished from that in crystalline solids \cite{sethna2017deformation} by the absence of well defined structural defects with which it can be associated. Thus, the structural aspects of plastic rearrangements \cite{Richard2020,Bonfanti2019,parmarprx19,DenisovSR15,VishwasPRE20,Mitra_2021} have been a subject of investigation, to understand the structural motifs associated with plastic rearrangements below yielding, and to investigate the structural features that distinguish the regions in which plasticity is concentrated. 

Another aspect of the approach to yielding and steady state flow that has received considerable attention is the distribution of {\it avalanches} corresponding to plastic rearrangements \cite{dahmenPRL09,LinPNAS14,regev2015reversibility,leishangthemNAT2017,oyama2020unified}, of interest also in a wide variety of phenomena exhibiting {\it crackling noise} \cite{Sethna2001}. The avalanche distribution is expected to have a power-law form, with a characteristic cutoff that is finite below yielding, with a mean field prediction of $\tau = 3/2$ for the power law exponent. The scaling form has been rationalised by several elasto-plastic models and mean-field theories constructed to pin down the scaling properties of avalanches \cite{LinPNAS14,JaglaPRE15,liu2016driving,BouchbinderPRE07,DahmenNATPHY2011,FranzPRB17}.
In numerical simulations, the avalanche distribution is found to be different across the yielding transition for cyclic shear \cite{leishangthemNAT2017}, and to depend on factors such as the inertia of the system \cite{salernoPRL12}, shear rate \cite{liu2016driving}, and the quantification of avalanche size (in terms of energy drops,  or the size of the connected clusters of active particles) \cite{leishangthemNAT2017}. The dependence of the characteristic size of the avalanches on system size have been analysed \cite{karmakarpre10,leishangthemNAT2017,ozawapnas2018}, with an observed $N^{1/3}$ scaling with the number of particles $N$. The implication of long range interactions on the break up of avalanches into clusters, and their statistics have been investigated for crack propagation \cite{Laurson2010,PriolPRL21}, but not,  to our knowledge, in the context of yielding of glasses. Performing such analysis, in addition to confirming key results in \cite{PriolPRL21}, we propose and verify a new relation between exponents characterising avalanches and clusters. 





Computational investigations of yielding in amorphous solids described above have largely been performed for solids with particles interacting with spherically symmetric, short ranged interactions. In particular, relatively few studies \cite{LeonfortePRL06,MantisiEPJE12,richardjpcc17,RountreePRL09,BonfantiNL18,Bonfanti2019} have addressed the archetypal glass, silica, which is characterised by an open, tetrahedral, local geometry, and whose interaction potential includes long range Coulomb interactions 
(or silicon \cite{DemkowiczPRB05liquid,DemkowiczPRB05avalanche,ArgonPM06,FuscoPRE10}, which shares several geometric and thermodynamic characteristics). In the liquid state, the tetrahedral network structure of silica entails a rich spectrum of novel behavior, including  density maxima \cite{angellSC76,stanleyLP}, a liquid-liquid phase transition \cite{voivodpre00,chenJCP17} and a strong-to-fragile transition \cite{horbachPRB99, hess1996parametrization, saksaengwijit2004origin,voivodPRE04a}. 
It is of interest to investigate the role of such directional, tetrahedral local geometry, and of long range interactions in the yielding behavior of silica and, in particular, the nature of avalanches  and the structural changes involved in plasticity and strain localisation. The yielding behavior of silica under cyclic shear has been shown to be broadly similar to that for the Kob-Andersen binary Lennard-Jones mixture (KA-BMLJ) 
\cite{BhaumikPNAS21}, characterised by a qualitative change across a threshold temperature of $T_{th} = 3100K$ (see Fig. S1 in Supplemental Material (SM) for illustration). In contrast, we show in this letter that the nature of avalanches and structural change associated with yielding 
display unusual features in the case of silica.

We study a  version of the BKS model introduced by Saika-Voivod \cite{beestPRL90,voivodPRE04a} (see SM for details). We prepared several equilibrated samples by performing constant temperature (NVT) molecular dynamics simulations with an integration time step of 1fs for a wide range of temperatures that straddles the threshold temperature $T_{th}=3100K$ \cite{BhaumikPNAS21} for a fixed density $\rho=2.8 ~g/cm^3$. Avalanche properties display significant size dependence and, for this reason,  we also simulate sizes ranging from $N=1728$ to $N=74088$. All the  samples are equilibrated for at least  $20\tau_\alpha$, $\tau_\alpha$ being the structural relaxation time obtained from the self intermediate scattering function $F_s(k,t)$. 
Inherent structures (energy minimum configurations) obtained from instantaneous quenches of equilibrated liquid configurations are then subjected to an  athermal quasi-static shearing (AQS) protocol involving two steps: (i) affine deformation by a small strain increments of $d\gamma = 2\times 10^{-4}$ in the $xz$-plane ($x^\prime\to x+d\gamma ~z$, $y^\prime\to y$, $z^\prime\to z$) and (ii) energy minimization. The procedure is then repeated and the strain $\gamma$ is varied cyclically as : $0\to \gamma_{\rm {max}}\to -\gamma_{\rm{max}}\to 0$. Repeating the deformation cycle for a fixed strain amplitude $\gamma_{\rm{max}}$, the glasses are driven to the steady state wherein properties of the system remain stable with further cycles of strain. We consider $12$ samples for $N=1728$, 4 samples for $N=5832$ and $13824$, and one sample for $N=27000$ and $N=74088$ to perform the cyclic shear. We employ the conjugate-gradient algorithm for energy minimization and execute all the numerical simulations in LAMMPS \cite{plimptonjcp1995}.



\begin{figure}[t]
  \centerline{
  \includegraphics[width=1\linewidth]{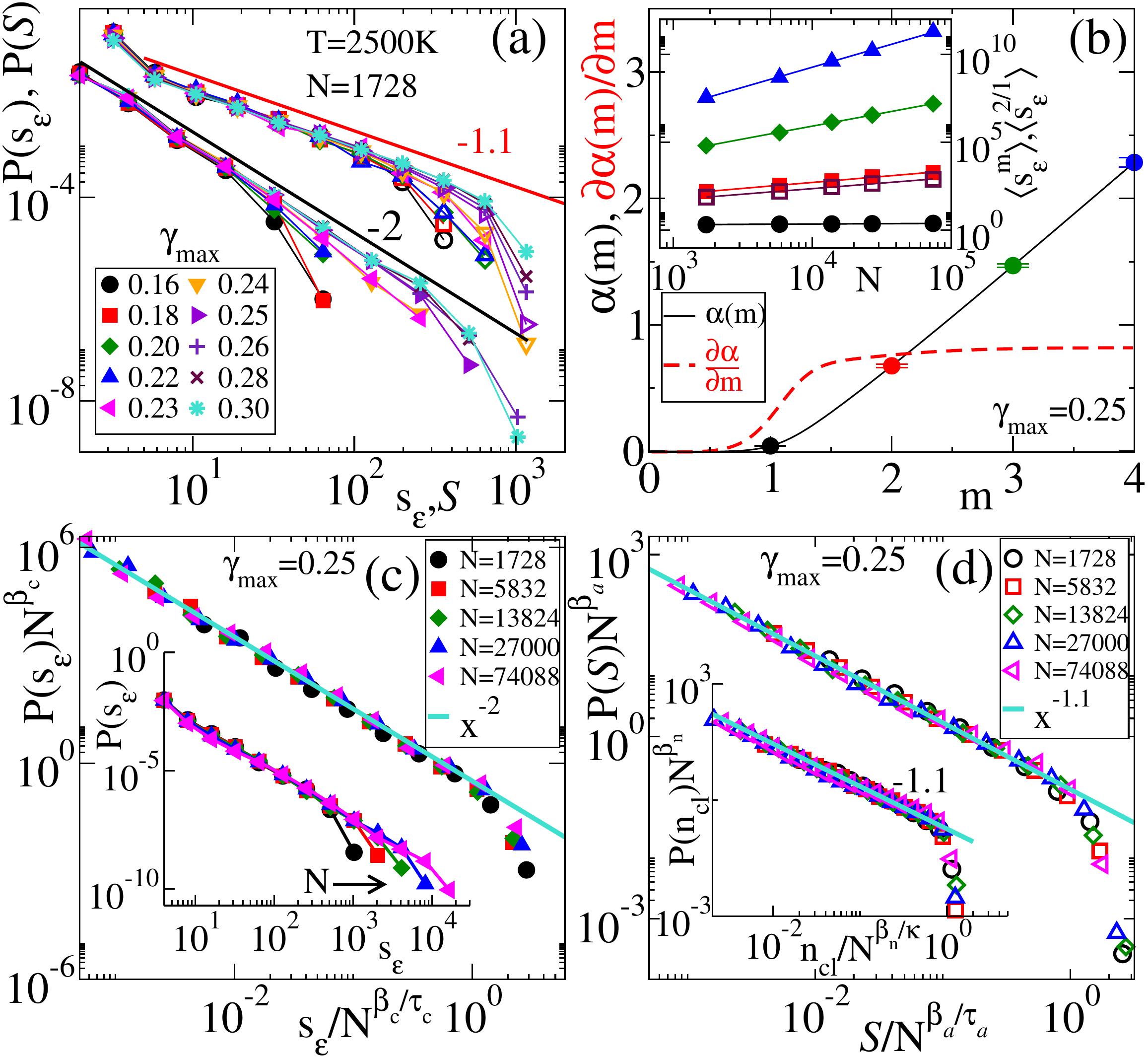}}
\caption{\label{Fig_avalanche} (a) Distribution of avalanches size (open symbols) and clusters size (filled symbols) of active particles for $T=2500K$ for several strain amplitudes $\gamma_{max}$ (The yield amplitude $\gamma_{max}^{Y} = 0.23$). (b) Moment analysis of cluster size: The moment exponent (see text) $\alpha(m) = (\beta_c/\tau_c)(m+1-\tau_c)$ (black line) and $\partial \alpha(m)/\partial m $ (red dashed line) against $m$. Points are highlighted for integer values of $m$ for which data of $\langle s_\epsilon^m \rangle$ against $N$ are shown in the inset. The solid lines in the inset are the least squares fits to extract the value of $\alpha(m)$. Open squares represent the ratio of the second and first moments $\langle s_{\epsilon}^{2/1}\rangle=\langle s_{\epsilon}^2\rangle / \langle s_{\epsilon} \rangle$ which scales as $N^{0.63}$. (c) $P(s_\epsilon)N^{\beta_c}$ against $s_\epsilon/N^{\beta_c/\tau_c}$ for different system size $N$ with $\tau_c=2.15$ and $\beta_c/\tau_c=0.79$. The solid line through the data points is a fit to $y \sim x^{-2}$. Inset shows the unscaled distributions for different system sizes. (d) Scaled avalanche size distribution $P(S)N^{\beta_a}$ for different system sizes with $\tau_a=1.1$ and $\beta_a/\tau_a=0.79$. Inset: Scaled distribution of the number of clusters for different system sizes with $\kappa=1.12$ and $\beta_n/\kappa=0.78$. Avalanches are collected in the first quadrant of the strain cycle.}
\end{figure}

We investigate avalanches by computing the statistics of {\it avalanche size} ($S$), {\it cluster size} ($s$), and the {\it number of clusters} ($n_{cl}$). The size of the avalanches is computed as the total number of {\it active} particles during a plastic rearrangement, identified by computing the deviatoric strain $\epsilon_d$ for each particle. Active particles are identified as those for which $\epsilon_d > 0.22$, following the procedure introduced in \cite{salernoPRE13} (see SM). We further obtain the sizes of clusters of connected active particles. Distributions of avalanche size and cluster size for several $\gamma_{max}$ are shown in Fig. \ref{Fig_avalanche}(a) and follow  power laws with exponents close to $\tau_a = 1.1$ for avalanches and $\tau_c = 2$ for clusters, with  $\gamma_{max}$ dependent cut-offs in each case. Strikingly, the cluster size exponent ($\tau_c$) is significantly greater than the mean field value,  $3/2$~\cite{sethna2017deformation}, whereas $\tau_a$ is significantly smaller.  
The distributions of energy drops, however, follow a  power-law with exponent $\approx -1.25$ (see SM) as also observed for the KA-BMLJ \cite{leishangthemNAT2017} for which $\tau_c = 3/2$.

In order to confirm these exponents, we perform a finite scaling analysis of the distributions of $S$, $s_{\epsilon}$ and $n_{cl}$, for $\gamma_{max} = 0.25$ (consistent results for other $\gamma_{max}$ are shown in the SM). We assume a scaling form for cluster size 
\begin{equation}
P(s_\epsilon)\approx N^{-\beta_c}f\left [s_\epsilon/N^{\beta_c/\tau_c}\right],
\end{equation}
where the scaling function $f(x)\to x^{-\tau_c}$ for $x \to 0$, and $f(x)\to 0$ for $x\to 1$. This scaling form implies that the moments $\langle s_\epsilon^m \rangle\sim N^{\alpha(m)}$, where
$\alpha(m)=\beta(m+1-\tau_c)/\tau_c$ is the moment exponent \cite{MenechPRE98,chessa1999universality} (see  SM for details). In the inset of Fig.~\ref{Fig_avalanche}(b), we show a log-log plot of $\langle s_\epsilon^m \rangle$ against $N$ for $m=1,2,3$, and $4$, from which we obtain $\alpha(m)$. In Fig. \ref{Fig_avalanche}(b), we present  $\alpha(m)$ and the corresponding derivative $\partial \alpha(m)/\partial m$ (which must equal $\beta_c/\tau_c$ for large $m$) as a function $m$. By a linear fit of $\alpha(m)$ in the large $m$ range, we determine $\beta_c/\tau_c=0.79\pm 0.02$ and $\beta_c=1.70\pm 0.10$. Fig. \ref{Fig_avalanche}(c) shows the scaled distributions $P(s_\epsilon,N)N^\beta_c$ plotted against the scaled variable $s_\epsilon/N^{\beta_c/\tau_c}$, using these values, for different system size $N$ to obtain the data collapse which supports the validity of the assumed scaling function. However, the  collapsed data is best described by $\tau_c = 2$ (close to, but slightly smaller than, $\tau_c = 2.15 \pm 0.07 $ obtained from $\beta_c/\tau_c$, $\beta_c$ above) which we treat as our estimate below (see SM, Fig. S7, that further supports the value $\tau_c = 2$). Assuming similar scaling forms for $S$ and $n_{cl}$, we estimate $\tau_a=1.1\pm 0.05$, $\beta_a/\tau_a=0.79\pm 0.02$ for avalanche size, and $\kappa=1.12\pm 0.08$, $\beta_n/\kappa=0.78 \pm0.03$ for number of clusters.
In Fig. \ref{Fig_avalanche}(d) we present the collapsed data for $S$ and $n_{cl}$ that confirm these exponents.

We next discuss the relationship between these exponents. Considering $n(s|S)$, the number of clusters of size $s$ in an avalanche of size $S$, we have, by defintion, $\int_{1}^{S} ~ s~ n(s|S) ~ds = S$ and $\int_{1}^{S} ~  n(s|S)~ ds = n_{cl}(S)$. We assume (as supported by numerical data) that $n(s|S) \sim s^\tau$ up to the cutoff $S$, but importantly, $\tau \ne \tau_c$. We straight-forwardly obtain (see SM for details) $\langle n_{cl}\rangle_S\sim S^{\gamma_{ns}  }$ with 
$\gamma_{ns} = \tau - 1$, and the mean cluster size $\langle s\rangle_S\sim S^{2 - \tau}$ (see also \cite{PriolPRL21}). As shown in the SM, numerically, we obtain $\langle n_{cl}\rangle_S\sim S^{0.9}$ and  $\langle s\rangle_S\sim S^{0.1}$, leading to $\tau = 1.9 \ne 2$. Further, assuming a scaling function $P(n_{cl}|S)\sim S^{-\gamma_{ns}}g(n_{cl}/ S^{\gamma_{ns}})$ for the distribution of the number of clusters, we obtain 
\begin{equation}
P(n_{cl})=\int P(n_{cl}|S)P(S)dS \sim n_{cl}^{-(1+(\tau_a-1)/\gamma_{ns})}, 
\label{pncl_n}
\end{equation}
or, $\kappa = 1 + (\tau_a-1)/\gamma_{ns} =  1 + (\tau_a-1)/(\tau - 1)$. Considering the normalised distribution $P(s|S)$ and writing 
\begin{equation}
P(s) = \int_{s}^{\infty} P(s|S) P(S) dS \sim s^{-(\tau + (\tau_a -1))},
\label{sdist}
\end{equation}
we obtain a new relation between the avalanche and cluster size exponents, 
\begin{equation}
\tau_c = \tau - 1 + \tau_a = \gamma_{ns} + \tau_a. 
\end{equation}
The exponent values we obtain, 
\begin{equation}
\tau = 1.9 ,\ \  \gamma_{ns} = 0.9  ,\ \   \tau_c = 2 ,\ \ \tau_a = 1.1 ,\ \  \kappa = 1.12
    \label{allexp}
\end{equation}
clearly satisfy the exponent relationships we describe. Such consistency is also obtained for a two dimensional glass (detailed in an accompanying paper \cite{bhaumik2D2021}). Despite such consistent analysis within the framework of \cite{PriolPRL21}, the large value of $\tau_c$ is surprising. Similar values have been discussed for silica nanofibres  \cite{BonfantiNL18}, amorphous silicon \cite{DemkowiczPRB05avalanche} and in preliminary results for a short ranged silica-like model \cite{CoslovichJPCM2009}. Although the details in these systems differ, we speculate that the open framework structure common to these systems may provide an explanation.

We also carry out the cluster analysis for BKS Silica by identifying {\it active} particles using non-affine displacements as reported in \cite{leishangthemNAT2017} for KA-BMLJ, and employ the deviatoric local strain for analysing the KA-BMLJ system. While the former analysis for silica yields $\tau_c=2$, we obtain $\tau_c=3/2$ for KA-BMLJ, in agreement with \cite{leishangthemNAT2017} (See  SM).  A summary of avalanche exponents found in different models is included in the SM. Finally, we note that $d \beta_c/\tau_c$ yields the fractal dimension $d_f^{est}$. The estimated $d^{est}_f = 2.37$ for silica ($d^{est}_f = 1.8$ for KA-BMLJ, for which we find $\beta_c/\tau_c = 0.6$), is close to the value obtained directly using the box counting method, $d_f = 2.22$ ($d_f = 2$ for KA-BMLJ) (see SM for details).

\begin{figure}[t]
  \centerline{
  \includegraphics[width=1\linewidth]{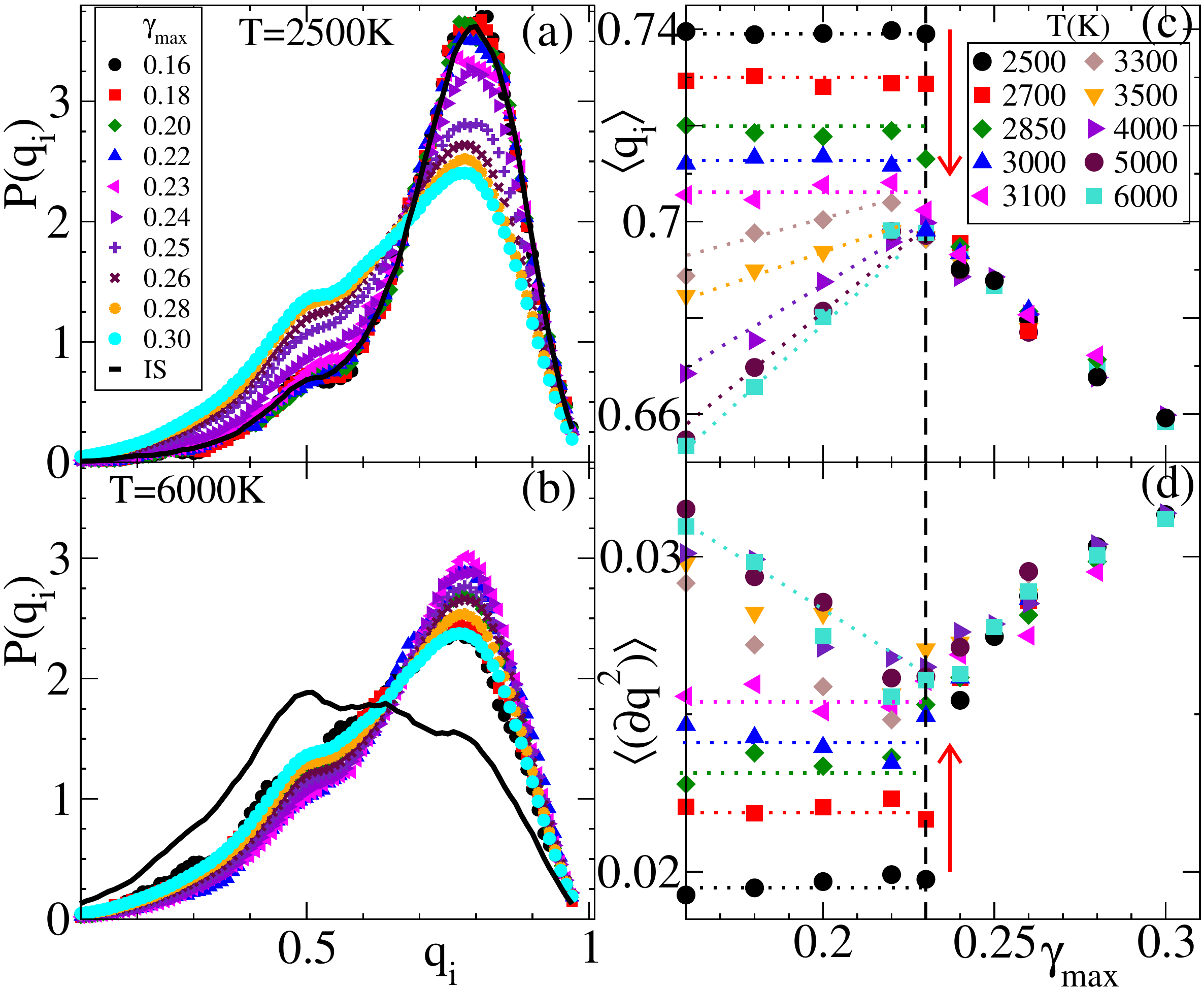}}
\caption{\label{qi_gmax} Distributions $P(q_i)$ of the tetrahedrality parameter for zero strain configurations of cyclically deformed silica for different strain amplitude $\gamma_{max}$ for (a) $T=2500K$ and (b) $T=6000K$. (c) Averages $\langle q_i \rangle$ and (d) Variances $\langle (\partial q)^2 \rangle = \langle q^2\rangle -\langle q \rangle ^2$, as a function of $\gamma_{max}$ for different temperatures $T$. Data are averaged over several configurations collected from different samples in the steady state for each $\gamma_{max}$. The vertical dashed line indicates the yield strain, dotted lines through data points are guide to the eyes and the arrows indicate the direction of increasing temperature.}
\end{figure}

We next study the modification of structure under shear by considering 
the tetrahedrality parameter \cite{shellpre02}, 
\begin{equation}
 q_i=1-\frac{3}{8}\sum_{j>k}\left[\cos \theta_{jik}+\frac{1}{3}\right]^2,   
\end{equation}
that  measures  the tendency of neighboring silicon atoms to form a tetrahedral structure around a central atom $i$ and is equal to unity for a perfectly tetrahedral local environment (see SM for details). We compute the distributions  $P(q_i)$ for different $\gamma_{max}$ across the yielding transition for two representative temperatures above and below the threshold  temperature $T_{th}$. The results are shown in Figs. \ref{qi_gmax}(a) and \ref{qi_gmax}(b) for $T=2500K$ and $6000K$, respectively. For $T=2500K$, the structure does not display any evolution below yielding and the distributions are indistinguishable for $\gamma_{max}<\gamma^Y_{max}$. The peak at $q\sim 0.8$ indicates a high silicon-silicon tetrahedral order that does not vary  with $\gamma_{max}$ ~\cite{shellpre02}, and the structure  is indistinguishable from that of the initial undeformed samples. Beyond yielding, the  distributions evolve and become broader with increasing amplitude $\gamma_{max}$. 
The high $q_i$ peak value becomes less pronounced and shoulder around $q_i\simeq 0.4$ appears. Interestingly, the variation of $P(q_i)$ with increasing $\gamma_{max}$ has a strong resemblance to what is observed in equilibrium when temperature is increased~\cite{shellpre02}. For $T=6000K$, Fig.~\ref{qi_gmax}(b), the behaviour is very different. Below yielding, a strong enhancement of the tetrahedral order upon increasing $\gamma_{max}$ is observed, with the initial undeformed glass displaying  very weak tetrahedral order. This effect  is reflected in the growth of the high $q_i$ peak  which continues until the yielding amplitude is reached. Beyond yield, similarly to what is observed for the low temperature case, the peak value decreases with $\gamma_{max}$.

To further characterize the structural disorder induced by deformation, we study the mean and variance of $q_i$ as a function of strain amplitude for different $T$ as shown in Figs.~\ref{qi_gmax}(c) and \ref{qi_gmax}(d)  respectively. 
As expected, two very different trends are observed below and above yielding. Below yielding, we observe again two patterns. For $T > T_{th}$, $\langle q_i\rangle$ progressively increases with $\gamma_{max}$ up to the the yield amplitude. Interestingly, the maximum orientational order is obtained at the yielding amplitude where, for all the cases  with $T\ge T_{th}$, it converges to $0.7$ which is the value of $\langle q_i\rangle$ for the undeformed samples at $T_{th}$ (see Fig. S16 of SM). For $T < T_{th}$, $\langle q_i\rangle$ does not vary with $\gamma_{max}$ until the yield point where it abruptly drops to  the same values as for the high temperature case. Above yielding, all the curves collapse, indicating that the final structure depends only on the strain amplitude and not on the initial temperature.  
In this regime,  $\langle q_i\rangle$  decreases with $\gamma_{max}$, indicating a progressively less tetrahedral structure. As shown in Fig.~\ref{qi_gmax}(c), the fluctuation of $q_i$ also behaves in a similar way but in an opposite fashion. Interestingly, at the yielding point the fluctuations are at their minimum. These trends strikingly reflect the changes in the energy of the system ~\cite{BhaumikPNAS21}. The decrease of tetrahedral order in deformed silica can be explicitly linked to an increase in the population of $5$-coordinated silicon atoms, as shown in Sec. S-7 of the SM.

\begin{figure}[!t]
     \centerline{
  \includegraphics[width=1\linewidth]{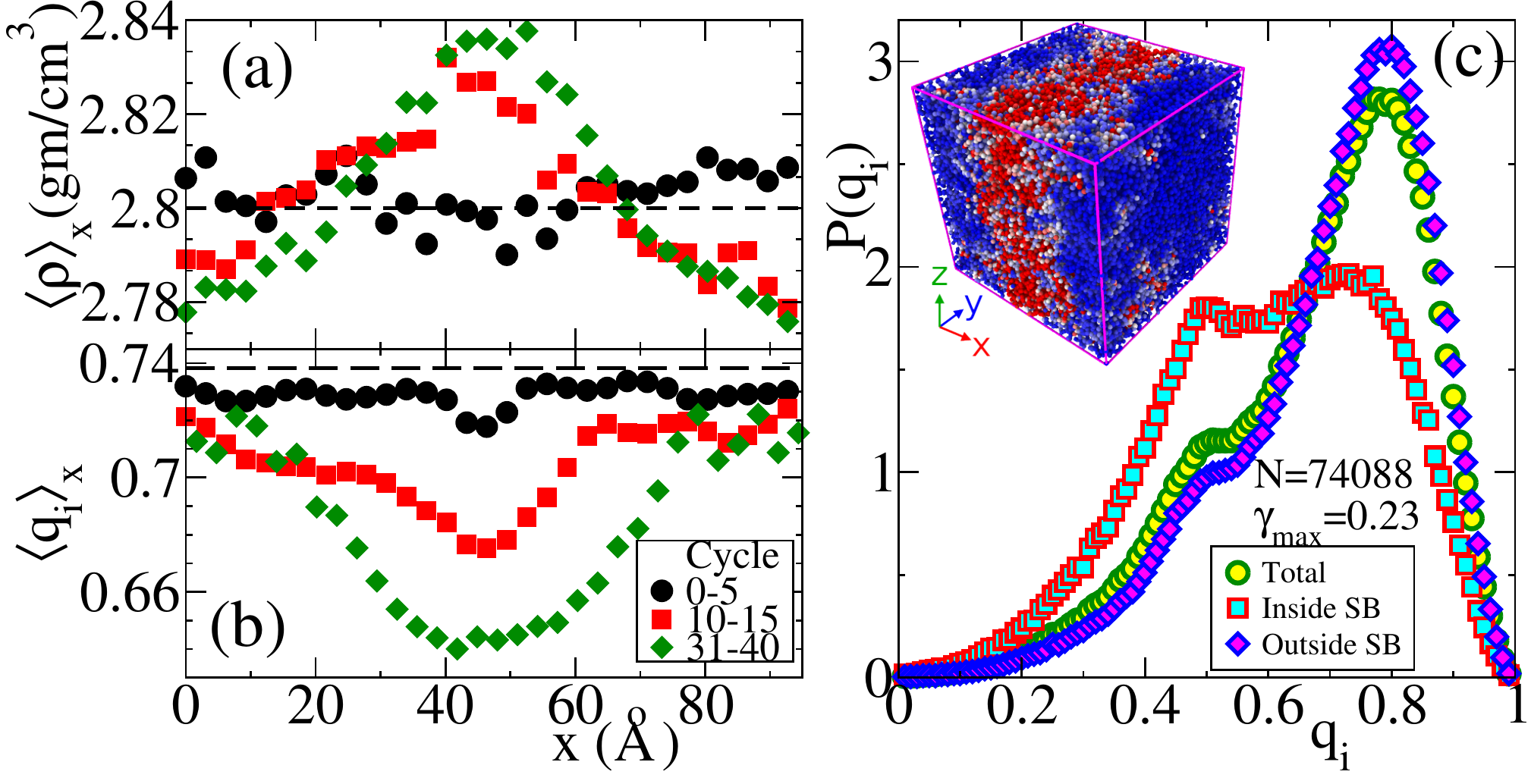}
}
\caption{\label{shearband} (a) Slab-wise averaged density $\langle \rho \rangle_x$ {\it vs.} coordinate $x$. The dashed horizontal line at $2.8 g/cm^3$ indicates the global density. (b) Slab-wise averaged $\langle q_i\rangle_x$ {\it vs.} $x$,  averaged over consecutive cycles at three different windows. The dashed horizontal line indicates the value of $q_i$ of the initial undeformed glass. (c) Distribution $P(q_i)$,  for the total system, and for atoms inside, and outside the shear band. To calculate the $q_i$ inside the shear band, we consider those atoms whose $\epsilon_d>1.25$. Inset: Snapshot of a steady state stroboscopic configuration for $\gamma_{max}=0.23$ for system size $N=74088$ and $T=2500K$. The color map indicates the deviatoric strain  $\epsilon_d$ between successive stroboscopic configurations. 
}
\end{figure}

We next investigate structural features associated with  strain localization ~\cite{parmarprx19} above the yield strain amplitude. In Fig.~\ref{shearband}(c) (inset), we show a snapshot of a zero strain configuration for the largest system size simulated ($N=74088$) at a strain amplitude $\gamma_{max}=0.23$. The color map corresponds to the deviatoric strain $\epsilon_d$, computed between two consecutive stroboscopic configurations, up to a cut-off value $1.25$ (See SM for the discussion about this choice), highlighting the localisation of strain in a shear band.  In Ref.~\onlinecite{parmarprx19,Mitra_2021} the density within the 
shear band was shown to be less than the average density. We compute and plot the slab-wise density $\langle \rho_x\rangle$ along $x$-direction in Fig.~\ref{shearband}(b). Contrary to 
the observation in \cite{parmarprx19,Mitra_2021}, we find that $\langle \rho_x \rangle$ becomes progressively larger inside the shear band, with the number of cycles of shear. 
This reversal of trend is clearly a reflection of the fact that the energetically favorable tetrahedral structure of silica has lower density than more disordered structures, which leads to well-known density and other anomalies in silica~\cite{shellpre02}. In order to verify this expectation, we compute slab-wise averages of $q_i$, which are shown in Fig.~\ref{shearband}(c). These results clearly demonstrate that the higher density structure within the shear band also has reduced orientational order, analogous to observations in ~\cite{ShiPRB06,VishwasPRE20}. The suppressed tetrahedral order within the shear band is associated with the enhancement of the fraction of $5$-coordinated defects  (See Fig. S19 of SM). We compute the distributions of $q_i$ within and outside the shear band, and compare with the aggregate distribution  in Fig.~\ref{shearband}(d). These distributions reveal the structure within the shear band to be comparable to high temperature undeformed glasses, whereas outside, the are comparable to low temperature glasses. 

In summary, we have investigated the statistics of avalanches and clusters in silica and obtained a satisfactory analysis of the relationship between exponents within a framework \cite{PriolPRL21} that envisages the fragmentation of avalanches in the presence of long range interactions. We have further proposed and verified a new relation between avalanche and cluster exponents. How the microscopic structure may lead to the fragmentation of avalanches is an interesting question to investigate further. 
We have also investigated structural change across the yielding transition and differences in structure within and outside shear bands and have found that yielding and the formation of shear bands is accompanied by a reduction of tetrahedral order, which corresponds to an anomalous increase (rather than decrease) of density. Although the qualitative features of yielding in silica are analogous to other glass formers, the special features of local geometry in silica apparently lead to unusual avalanche statistics and structural change during yielding. 

\paragraph*{Acknowledgements.} 
We thank J. Horbach, S. Zapperi and A. Rosso for useful discussions and comments on the manuscript. We acknowledge Indo-French Center for the Promotion of Advanced Research (IFCPAR/CEFIPRA) Project 5704-1 for support, the Thematic Unit of Excellence on Computational Materials Science, and the National Supercomputing Mission facility (Param Yukti) at the Jawaharlal Nehru Center for Advanced Scientific Research for computational resources. S.S. acknowledges support through the J. C. Bose Fellowship  (JBR/2020/000015)  SERB, DST (India).

\bibliographystyle{apsrev4-1} 
\bibliography{ref}

\end{document}